\documentclass[12pt,preprint]{aastex}
\def\ra#1#2#3{#1$^{\rm h}$#2$^{\rm m}$#3$^{\rm s}$}
\def\dec#1#2#3{$#1^\circ#2'#3''$}

\def\swift{{\it Swift}}
\def\sst{{\it Spitzer}}
\def\um{{$\mu$m}}

\def\ociw{2}
\def\prince{3}
\def\hubble{4}
\def\ssc{1}
\def\uh{5}

\slugcomment{Submitted to {\it The Astrophysical Journal}}
\shorttitle{{\it Spitzer} Observations of GRB Hosts}
\shortauthors{Chary et al.}

\begin{document}

\title{\sst\ Observations of Gamma-Ray Burst Host Galaxies: A Unique Window into High Redshift Chemical Evolution and Star-formation}

\author{
R.~Chary\altaffilmark{\ssc},
E.~Berger\altaffilmark{\ociw,}\altaffilmark{\prince,}\altaffilmark{\hubble},
L.~Cowie\altaffilmark{\uh}
}

\altaffiltext{\ssc}{Spitzer Science Center, California Institute 
of Technology, Mail Stop 220-6, Pasadena, CA 91125}

\altaffiltext{\ociw}{Observatories of the Carnegie Institution
of Washington, 813 Santa Barbara Street, Pasadena, CA 91101}
 
\altaffiltext{\prince}{Princeton University Observatory,
Peyton Hall, Ivy Lane, Princeton, NJ 08544}
 
\altaffiltext{\hubble}{Hubble Fellow}

\altaffiltext{\uh}{Institute for Astronomy, University of Hawaii, HI 96822}

\begin{abstract} 

We present deep 3.6\,\um\ observations of three $z\sim 5$ GRB host
galaxies with the {\it Spitzer Space Telescope}.  The host of
GRB\,060510B, at $z=4.942$, is detected with a flux density of
$0.23\pm 0.04$ $\mu$Jy, corresponding to a rest-frame $V-$band
luminosity of $1.3\times 10^{10}$ L$_{\sun}$, or $\approx 0.15$ L$_{*, {\rm V}, z=3}$.
We do not detect the hosts of GRBs\,060223A and 060522 and constrain
their rest-frame $V-$band luminosity to $<0.1$ L$_{*, {\rm V}, z=3}$.  Our
observations reveal that $z\sim 5$ GRB host galaxies are a factor of
$\sim 3$ less luminous than the median luminosity of
spectroscopically-confirmed $z\sim 5$ galaxies in the Great
Observatories Origins Deep Survey (GOODS) and the {\it Hubble} Ultra Deep Field (UDF).
The strong connection between GRBs and massive star formation implies
that not all star-forming galaxies at these redshifts are currently being
accounted for in deep surveys and GRBs provide a unique way to measure
the contribution to the star-formation rate density from galaxies at
the faint end of the galaxy luminosity function.  By correlating the
co-moving star-formation rate density with co-moving GRB rates at
lower redshifts, we estimate a lower limit to the
star-formation rate density of $0.12\pm 0.09$ and $0.09\pm 0.05$
M$_{\sun}$ yr$^{-1}$ Mpc$^{-3}$ at $z\sim 4.5$ and $z\sim 6$,
respectively.  This is in excellent agreement with extinction corrected estimates from
Lyman-break galaxy samples.  Finally, our observations provide initial
evidence that the metallicity of star-forming galaxies evolve more slowly than the stellar mass density 
between $z\sim5$ and $z\sim0$, probably indicative of the loss of a significant
fraction of metals to the intergalactic medium, especially in low-mass galaxies.

\end{abstract}
 

\section{Introduction}
\label{sec:intro}

Our ability to measure the star-formation rate density (SFRD) at $z>4$
relies almost entirely on either narrow-band surveys which detect
strong Ly$\alpha$ emitting galaxies \citep{Hu:04,Nagao:07} or deep
imaging surveys of UV bright Lyman-break galaxies
\citep{Gia:04,Bouw:06,Hu:06}.  These surveys, by virtue of being flux
limited, trace the bright end of the galaxy luminosity function down
to $\sim 0.04$ L$_{*,{\rm UV},{\rm z=3}}$.  The various observations have
revealed a decline in the SFRD by about
a factor of 3 between $z\sim 3$ and $z\sim 6$
\citep{Bouw:06,Bunker:04}, with much of this decline being due to the
evolution of the bright end of the galaxy luminosity function.  

More than 
$90\%$ of the estimated SFRD at these redshifts takes place in
sub-L$_{*,{\rm UV},{\rm z=3}}$ galaxies.
In addition, spectroscopic confirmation of high redshift galaxies
relies on Ly$\alpha$ emission, which is easily obscured by dust.
There is now increasing evidence for rapid dust production within
$\sim 1$ Gyr of the Big Bang \citep{Chary:07, CSE:05, Maiolino:04}.  As a
result, quantifying possible dust extinction corrections and measuring
the faint-end slope of the galaxy luminosity function is essential for
minimizing uncertainties in the high redshift SFRD.

Measurements of the metallicity in typical star-forming
galaxies at $z>4$ is beyond the technological capability of the
current generation of instrumentation.  The relevant rest-frame optical
emission lines are very weak and are redshifted to the mid-IR.  The
alternative approach of studying chemical enrichment through damped
Ly$\alpha$ absorbers (DLAs) detected against background quasars
appears to be limited to $z\lesssim 5$ \citep{pgw:03, Songaila:02}, and is biased
towards tracing the properties of extended halo gas, which at lower
redshift, significantly underestimates the disk metallicity.  As a
result, the apparent evolution in the mass-metallicity and
luminosity-metallicity relations ($M$-$Z$ and $L$-$Z$) from $z=0$ to
$z\sim 2$ \citep[e.g.][]{Tremonti:04, KK:04, Savaglio:05} cannot be
traced to $z>5$, where such information should shed light on the
initial stages of mass build-up and metal enrichment in galaxies.

Long duration GRBs, by virtue of being associated with the deaths of
massive stars, provide a complementary technique for measuring the
SFRD and the chemical enrichment history.  \swift\ has revolutionized
this study by detecting GRBs out to $z\sim 6$ \citep{Gehrels:04, Kawai:06}.  Prompt spectroscopy of
the bright afterglows has now provided a sample of $\sim 20$ GRBs over
a wide range of redshifts with a wealth of metal absorption features
arising in the host galaxy \citep[e.g][]{Jensen:01, Castro:03, Hjorth:03, Vree:04, Fynbo:06}.  These observations provide a unique
window into the metallicity and gas column density in star forming
environments at high redshifts \citep{Chen:05, Berger:06, PCB:06, Price, Prochaska:07}.  Once
the afterglows fade away, deep observations of the field can also
reveal the stellar mass and star-formation rate of the host galaxies,
which can then be correlated with the inferred metallicities.

In order to 
study the host galaxies of high redshift GRBs and
take advantage of the diagnostics afforded by GRBs, we 
present {\it Spitzer Space Telescope} 3.6 \um\ observations of the
hosts of three GRBs at $4\lesssim z\lesssim 5$. Building on the
constraints provided by \citet{Berger:07}
on the host galaxy of GRB\,050904 at $z=6.295$, we discuss the nature of the host galaxies,
redshift evolution of the luminosity-metallicity relation and provide
an independent measure of the high redshift SFRD for comparison with
estimates from Lyman-break galaxy samples. Throughout this paper, we
adopt a $\Omega_{M}=0.27$, $\Omega_{\Lambda}=0.73$, H$_{0}$=71 km~s$^{-1}$~Mpc$^{-1}$
cosmology.

\section{Observations}
\label{sec:obs}

As part of \sst\ program GO\,20000 (PI: Berger) we observed the fields
of GRBs 060223A ($z=4.406$; Berger et al., in prep.), 060510B \citep[$z=4.942$;][]{Price}, and 060522
($z=5.110$; Berger et al., in prep.) with the Infrared Array Camera \citep[IRAC;][]{Fazio:04}
in the bandpasses centered at 3.6 and 5.8 $\mu$m (Table 1). The
observations were undertaken between September and November 2006,
after the afterglows associated with the GRBs had faded below the
detectability threshold.  As shown in Table 1, the GRB fields lie in
regions with ``low" to ``medium''-level zodiacal background and cirrus
of 13$-$28 MJy/sr at 24 $\mu$m on the date of the observations.  We
used 100 s integrations with about 130 medium scale dithers from the
random cycling pattern for total on-source integration times of $\sim
13000$ s at each passband.  The nominal $3\sigma$ point source
sensitivity limits are 0.26 and 2.4 $\mu$Jy, respectively.
 
Starting with the S14.4.0 pipeline-processed basic calibrated data
(BCD) sets we corrected the individual frames for muxbleed and column
pull down using software developed for the Great Observatories Origins
Deep Survey (GOODS).  Due to the presence of bright stars in the
field, many of the frames at 3.6 $\mu$m also showed evidence for
``muxstriping''.  This was removed using an additive correction on a
column by column basis.  The processed BCD frames were then mosaiced
together using the MOPEX routine \citep{mak:05} and drizzled onto a
$0.6\arcsec$ grid.  Astrometry was performed with respect to the
brightest 2MASS stars in the field which showed a peak-to-peak
astrometric uncertainty of $0.2\arcsec$ at 3.6 $\mu$m.

The location of the GRB hosts was determined by aligning the {\it
Spitzer} images against images of the afterglow from the \swift\
UV/optical telescope (060223A and 060522) and the Gemini Multi-Object
Spectrograph on the Gemini-north 8-m telescope (060510B).  For the
latter, the astrometric uncertainty is $0.09\arcsec$ in each
coordinate, while using the UVOT images we obtain an astrometric
uncertainty of about $0.6\arcsec$.

All three GRB locations show the presence of nearby ($3\arcsec$)
brighter galaxies: GRB\,060223A has two sources with flux densities of
7.2 and 9.1 $\mu$Jy at distances of $1.9\arcsec$ and $2.3\arcsec$ from
the GRB position; GRB\,060510B has a source with a flux of 6.1 $\mu$Jy
about 3.1$\arcsec$ from the GRB position; and GRB\,060522 has a source
with 0.72 $\mu$Jy located 1.6$\arcsec$ away from the burst position.
This is not unexpected given the high source density in deep IRAC images.  
%% Based on the source densities in deep GOODS
%% IRAC images (Dickinson et al., in prep.), we estimate that the probability of a 6 $\mu$Jy source falling
%% wihin a 3$\arcsec$ beam is 0.08 while there is a 25\% chance of a 0.7
%% $\mu$Jy source within the same area.  
We subtracted the contribution
of these sources, in order to obtain the strongest possible constraints
on the flux from the GRB host galaxies.

Photometry at the position of the host galaxies was performed in fixed
circular apertures of $1.2\arcsec$ radius with appropriate beam size
corrections applied as stated in the \sst\ Observer's Manual.  We
clearly detect a galaxy coincident with the position of GRB\,060510B
with a flux density of $0.23\pm 0.04$ $\mu$Jy at 3.6 $\mu$m, and a
$3\sigma$ upper limit of 2.4 $\mu$Jy at 5.8 $\mu$m (Figure 1).  For
GRBs 060223A and 060522, due to blending from nearby brighter sources
and the residual effects from muxbleed, we are only able to provide
$3\sigma$ upper limits to the flux of the host galaxy (Table 1).

\section{Luminosity and Metallicity of GRB Hosts}
\label{sec:prop}

Of the 4 GRB host galaxies at $z\sim 5$ observed in this program at 3.6 and 5.8 \um,
(including GRB\,050904), only
GRB\,060510B is clearly detected.  The observed 3.6 \um\ flux
densities/limits for these galaxies correspond to rest-frame $V-$band
luminosities of $\sim 0.15$ L$_{*,{\rm V},{\rm z=3}}$, where L$_{*,{\rm V},{\rm z=3}}$ is
about $8\times 10^{10}$ L$_{\sun}$ \citep{mar:07, Shapley:01}. 
It is illustrative to compare the properties of GRB hosts with 
the field galaxy population at similar redshifts. 

The GOODS fields have extensive spectroscopy of galaxies at high redshift
\citep{Vanzella:05, Eros:06, Eros:07}.
There are 275 Lyman-break galaxies in both the GOODS fields 
which are classified as $V-$band ``dropouts'' i.e. $z\sim5$.
The magnitude limit of the GOODS optical observations imply that they are
brighter than $\sim 0.2$ L$_{*,{\rm UV},{\rm z=3}}$ \citep{Gia:04},
Of these, $\sim$20\% have spectroscopic redshifts while $\sim$30\% are
individually detected with IRAC. At higher redshifts, $z\sim6$,
it has been shown that galaxies which are individually undetected with IRAC 
appear to harbor a younger stellar population and have a factor of 10 lower
stellar mass than IRAC detected galaxies \citep{Yan}.

As shown in Figure 2, GRB host galaxies are factors of $2-3$ times
fainter than the median $V-$band luminosity of 
galaxies which have spectroscopic redshifts of $4.5<z<5.5$ in the GOODS field.
Furthermore, the luminosities are comparable to the rest-frame $V-$band luminosity of 
GRB hosts studied at lower redshifts \citep[e.g.;][]{CBA:02, lef:03}.
This suggests that GRB host galaxies are unlike the luminous end of the star-forming, Lyman-break
galaxy population which have had about a factor of 10 increase in their stellar mass
between $z\sim5$ and $z\sim1$. They are more typical of the blue, faint end of the
galaxy $V-$band luminosity function, a population for which it is difficult to measure 
redshifts or metallicities, in the absence of GRBs, due to their inherent faintness. 

GRB host galaxies at $z\sim 0.5-3$, which have extensive multi-wavelength data, 
show clear evidence for very high specific star
formation rates indicating an on-going starburst \citep{CBA:02, LC:04,
Castro:06}.  We do not yet have constraints on the star formation
rates in the $z\sim 5$ host galaxies presented here, due to their intrinsic faintness in the
rest-frame UV \citep[see e.g.][]{Fruchter, Jakobsson:05}.  However, spectroscopy of the afterglows by
\citet{Price} and Berger et al. (in prep.) has revealed a
wealth of absorption lines which have been used to derive the
metallicity and gas column density in the vicinity of the burst.  

Absorption spectroscopy of the three bursts presented here have yielded
neutral hydrogen gas densities in their host galaxies of: ${\rm log}[N({\rm
HI})]=21.6\pm 0.1$ (060223A), ${\rm log}[N({\rm HI})]=21.3\pm 0.1$
(060510B), and ${\rm log}[N({\rm HI})]=21.0\pm 0.3$ (060522).  Thus,
all three systems are clearly DLAs, with column densities near the median of
the distribution for GRB-DLAs \citep{Berger:06, Jakob:06}.  In addition,
the metallicities of the GRB 060223A and 060510B systems have been determined from
the detection of weak metal lines.  For GRB\,060522 the
signal-to-noise of the spectrum is too low to clearly identify any
metal lines and an estimate of the metallicity is thus not possible.
In the case of GRB\,060223A, we find an upper limit on the column
density of \ion{S}{2} of ${\rm log}[N({\rm SiII})]<15.3$, leading to a metallicity
of ${\rm [S/H]}<-1.45$.  The non-detection of \ion{Fe}{2}$\lambda
1608$ leads to a limit of ${\rm [Fe/H]}<-2.65$, but we stress that
iron can be heavily depleted onto dust grains.  From the
detection of the \ion{Si}{2}$\lambda 1304$ line we find 
${\rm log}[N({\rm SiII})] \approx 15.3$, and hence ${\rm [Si/H]}\approx -1.8$.  As in the
case of iron, silicon is also strongly depleted, so we conclude that
the metallicity of the GRB\,060223A DLA is in the range of $\sim -1.8$
to $\sim -1.4$.  For GRB\,060510B, we use the
\ion{S}{2}$\lambda\lambda 1250,1253$ lines to measure ${\rm
log}[N({\rm SiII})]=15.6\pm 0.1$, and hence a metallicity, ${\rm
[S/H]}=-0.85\pm 0.15$ \citep[see also][]{Price}. 

The metallicity estimates of the GRB hosts along with their rest-frame $B-$band 
luminosities (assuming a $B-V$ color of 0, typical of star-forming galaxies) are
shown in Figure 3. 
Also shown for comparison are the metallicity-luminosity relationships for different
samples of field galaxies. 
Despite the one detection and two limits for the luminosity
of the host galaxies, the figure shows that the redshift evolution of metallicity 
at a fixed $B-$band luminosity that is seen between $0<z<2$, clearly extends out 
to $z\sim5$. 

The chemical enrichment of galaxies is directly related to their past history of star-formation since
supernovae and stellar winds are responsible for recycling the products of nucleosynthesis back into
the interstellar medium.
The stellar mass density is the time integral of the star-formation history.
By comparing the redshift evolution of the stellar mass density
($\rho_{*}$) \citep[e.g][]{Dickinson, Yan, Stark, Chary:07}, with the redshift evolution of the
metallicity ($Z$), we can search for evolution of the stellar initial mass function and assess the
role of feedback in the build-up of galaxies. The {\it Spitzer} observations of the hosts
are crucial, since they enable metallicity comparisons to be made at a fixed rest-frame
$V-$band luminosity, over a wide range of redshifts.

Due to the fact that we have constraints on the $V-$band luminosity and metallicity of 
only one $z\sim5$ GRB host, 
we make the assumption that the median metallicity at each redshift, is that of a galaxy which has
a similar luminosity as the GRB host. This is not an unreasonable assumption. Within the
observational uncertainties, the slope of
the mass-metallicity relation appears to be invariant between $z\sim0$ and $z\sim2$ \citep{Erb}. 
The metallicity values are obtained by effectively making a vertical cut
at $-20.8$ mag in Figure 3 and are determined to be $-0.85\pm$0.15, $-0.35\pm$0.1 and $0.33\pm$0.1 dex 
at redshifts of 5, 2.3 and 0 respectively. 
The average estimated $\rho_{*}$ at these redshifts are 1.4, 6 and 56 in units of 10$^{7}$~M$_{\sun}$~Mpc$^{-3}$
(See references above).
We performed a Monte-Carlo analysis to obtain the best fit between $Z(z)$ and $\rho_{*}(z)$.

Star-forming galaxies which fall on the local mass-metallicity relationship, show a scatter 
of $\sim$0.1 dex at bright luminosities and $\sim$0.2 dex at faint luminosities \citep{Tremonti:04}.
We use a random number generator to offset the stellar mass density and metallicity by the 
observed scatter from the mean values quoted above \citep[See][Table 3 for the range in stellar mass density]{Dickinson}. 
We fit for the relation between $Z(z)-\rho_{*}(z)$ and repeat the process 10000 times. 
We find that 
${\rm d}Z/d\rho_{*}$ appears to be invariant between $0<z<5$ and that $Z(z)\propto\rho_{*}(z)^{0.69\pm0.17}$. This
suggests that the chemical enrichment of star-forming
galaxies takes place at a slower rate than the build up of stellar mass.
This is presumably due to the loss of metals from low-mass galaxies by outflows
and stellar winds, an effect which is primarily responsible for the
mass-metallicity relation seen in the local Universe \citep{Tremonti:04} and $z\sim2$ Lyman-break 
galaxies \citep{Erb}. However, alternate mechanisms such as depletion of metals onto dust grains
cannot be ruled out at this time.

There is the possibility of a selection effect in this analysis. If long duration GRBs arise in collapsars,
they might preferentially be in low-metallicity galaxies. As a result, it is possible that
GRB hosts have a lower metallicity than the average field galaxy of the same rest-frame optical
luminosity. Although
GRB hosts appear to be have low luminosities in the rest-frame UV and $V-$band, the observational evidence does not
indicate that the hosts have an unusually low-metallicity for their luminosity. 
Metallicity of 
GRB host galaxies appear to span the 
range 0.1$-$1~Z$_{{\rm solar}}$ \citep[e.g][]{Berger:07b, Berger:06, Prochaska:07a} and some of the hosts have even been found
to be associated with dusty, infrared luminous galaxies \citep[e.g][]{lef:06}. 

Nevertheless, we assess the reliability of our derived
$Z(z)-\rho_{*}(z)$ relation by considering a bias in the metallicity of GRB environments.
If we assume that the metallicity of the GRB environment is higher by $>$0.3 dex compared to the mean metallicity
of a galaxy at its luminosity, it implies that the mean metallicity at $z\sim5$ for a field galaxy at the luminosity
of the GRB host is -1.15$\pm$0.15.
The best fit relation to the three points is then consistent with an exponent 
of unity i.e. $Z(z)\propto\rho_{*}(z)^{0.85\pm0.19}$ but has a worse $\chi^{2}$.
The corollary is that if GRB hosts were biased by 0.3 dex towards lower metallicities, compared to the mean metallicity
of a galaxy at its luminosity, the best fit relation is $Z(z)\propto\rho_{*}(z)^{0.52\pm0.16}$ which is a larger
deviation from unity.  
Furthermore, if there were a bias
in GRB host metallicities, the slope of the $Z(z)-\rho_{*}(z)$ relation derived above at a fixed
$B-$band luminosity, would have
a different value between $2<z<5$ and $0<z<2$ due to the fact that the $0<z<2$ relation is determined
from star-forming galaxies while the $2<z<5$ relation is derived from GRB hosts and Lyman-break galaxies.
This is inconsistent with our fits, although larger samples of GRB hosts are needed to eliminate suggestions
of bias.

Detection of individual GRB hosts at
high redshifts is likely to remain difficult, due to their intrinsic faintness. There is a clear need
for homogeneous infrared surveys of GRB host galaxies which will
enable stacking to be performed as a function of metallicity, gas
density and rest-frame ultraviolet properties. Within our sample,  
GRB050904 is a marginal IRAC detection \citep{Berger:07},
while GRB060223A is dominated by detector systematics. As a result,
we are unable to provide additional constraints using stacking.
Observations of a larger sample of GRB hosts, such as those currently being
targeted in \sst\ program GO4-40599 (PI: Chary),
will allow
the luminosity-metallicity relation to be measured at high redshift and
lead to a better understanding of the faint end of the galaxy luminosity
function, a regime which is currently inaccessible even through
ultradeep surveys like GOODS and the UDF.

\section{Evolution of the Star-formation Rate Density}
\label{sec:conc}

It is now well known from various mid-infrared, far-infrared and 
submillimeter surveys, that the star-formation rate density
at $z\sim0.5-2.5$ is dominated by infrared luminous galaxies 
with L$_{{\rm IR}}$=L$(8-1000\,\mu {\rm m})>10^{11}$~L$_{\sun}$ and L$_{{\rm IR}}$/L$_{{\rm UV}}\sim10-100$
\citep[e.g][]{tak:05, Denis, Chary:01}. At $z\gtrsim3$, current long-wavelength surveys, 
due to their limited sensitivity, are unable to detect
galaxies which harbor the bulk of the star-formation. Thus, rest-frame ultraviolet
observations of galaxies are the only avenue for probing star-formation at high redshifts.

The primary uncertainties
associated with quantifying the SFRD at $z>3$, are the contribution from galaxies
at the faint end of the UV-luminosity function and dust corrections.
Since sub-L$_{*, {\rm UV},z=3}$ galaxies contribute $\sim$90\% of the SFRD, measurement
of the faint-end slope of the ultraviolet luminosity function, where completeness 
corrections and surface brightness
dimming issues are significant, needs to be undertaken carefully \citep{Steidel, Bouw:06}.
Similarly, if extinction were a significant issue, the galaxies that dominate the
star-formation rate density would be UV-faint or undetected in magnitude limited
rest-frame ultraviolet surveys. GRBs are relatively insensitive to these limitations.
If the GRB rate density were correlated with the co-moving star-formation rate density
at lower redshifts, where cross-calibration between the UV and IR are in broad agreement,
measurement of the GRB rate density at $z>3$ could provide an independent pathway to
quantifying the SFRD \citep[see also e.g.;][]{Price:06}.  

The three parameters which are most likely to dominate the calibration between GRBs and the
star-formation rate density are the evolution of metallicity with redshift, evolution
of the initial mass function of stars and identification and spectroscopic follow up of the GRB afterglow. 
If long duration GRBs were to preferentially occur in low metallicity environments,
the increase in the average metallicity of the Universe with decreasing redshift would result in a higher
SFR/GRB-rate ratio at low redshift. Similarly, evolution of the stellar initial mass function
from a ``top-heavy" to a Salpeter mass function with decreasing redshift would increase the
SFR/GRB-rate ratio at low redshift. 
On the other hand, the
detection efficiency and spectroscopic completeness of GRBs should be increasing
with decreasing redshift, implying a lower SFR/GRB-rate ratio at low redshift.

Calibrating each of these parameters individually is challenging at the present time,
partly because the relationship between GRB rate and environment is not well known and
due to the fact that observational selection effects cannot be quantified.
Therefore, we need to rely on empirical comparisons between known star-formation rate
estimates and GRB rate densities to assess GRBs as a star-formation rate indicator.
This empirical comparison can be optimally done at $z<3$ since in this
redshift range the star-formation rate, including the dust obscured component,
has been accurately determined from deep mid-infrared and submillimeter
surveys. 

We use the star-formation rate density at $z<3$ from
\citet{Chary:01}.  We distribute the 52 \swift\ GRBs with spectroscopic
redshifts into redshift bins
and divide by the co-moving volume in each redshift bin. We also
correct for the time dilation to estimate the comoving GRB rate
density over the $\sim 2$ year \swift\ lifetime.  The redshift bin at
$z<0.5$ is omitted since the GRB rate density appears to be anomalously
high compared to the rapidly evolving star-formation rate density.
We find that within the uncertainties, the
rate density of GRBs with spectroscopic redshifts in the range
$0.5<z<3$ is constant at a value of $(3.7\pm 1.1)\times 10^{-11}$
Mpc$^{-3}$ yr$^{-1}$.  This can be compared with the extinction-corrected
comoving
star-formation rate density in the same redshift range which is 
in the range 0.12$-$0.25\,M$_{\sun}$ yr$^{-1}$ Mpc$^{-3}$
and has an average value of $\sim 0.2$ M$_{\sun}$ yr$^{-1}$ Mpc$^{-3}$
\citep{Chary:01}.

Since these two independent rate densities are relatively constant in the 
$0.5<z<3$ range, we can tentatively make the assumption that the SFR/GRB-rate
is constant (Figure 4). The ratio of these two rates implies:
\begin{equation}
{\rm SFRD} = {\rm GRB~rate} \times (5.2\pm 2.3)\times10^{9},
\end{equation} 
where SFRD is the extinction-corrected star-formation rate density in
M$_{\sun}$ yr$^{-1}$ Mpc$^{-3}$ and GRB rate is in units of Mpc$^{-3}$
yr$^{-1}$.  

Using our derived calibration, and the measured GRB rate densities at
$4<z<5$ and $5<z<7$ of $(2.4\pm 1.2)\times10^{-11}$ and $(1.8\pm
0.9)\times10^{-11}$ Mpc$^{-3}$ yr$^{-1}$, respectively, we infer a net
star-formation rate, corrected for extinction, of $0.12\pm 0.06$ and
$0.09\pm 0.05$ M$_{\sun}$ yr$^{-1}$ Mpc$^{-3}$ at $z\sim 4.5$ and
$z\sim 6$ respectively. These estimates are systematically higher
than those derived by \citet{Price:06} by factors of $3-5$. The \citet{Price:06} estimates
were calibrated at $z\sim3$ where neither the completeness correction
factor for the faint end of the UV luminosity function nor the dust
extinction correction are reliably known while deep {\it Spitzer} mid-infrared
surveys have confirmed the dominant contribution of infrared luminous galaxies
to the star-formation rate density at $0.5<z<3$ \citep{tak:05, Daddi:07}.
The fact that the GRB rate density
is almost flat between $0.5<z<6$, while parameters such as the
detection efficiency and spectroscopic completeness should be decreasing
with increasing redshift, implies that the measured
GRB rate density provides at least a
lower limit to the star-formation rate density.

It is illustrative to compare this star formation rate estimate with
those from deep rest-frame ultraviolet surveys at
$z>4$. \citet{Gia:04} derive a star-formation rate density at $z\sim
4$ of 0.02 M$_{\sun}$ yr$^{-1}$ Mpc$^{-3}$ when integrating to 0.2
$L_{*,{\rm UV},{\rm z=3}}$.  After application of an extinction correction of
$A_{\rm V}=0.45$ mag, based on the extinction properties in local
starburst galaxies, they estimate the total star-formation rate
density at $z\sim 4$ to be 0.15 M$_{\sun}$ yr$^{-1}$ Mpc$^{-3}$.
Similarly, \citet{Bouw:06} derive a star-formation rate density at
$z\sim 6$ by integrating the luminosity function of Lyman-break
galaxies in the UDF and other deep fields. Integrating the UV luminosity function
down to 0.2\,$L_{*, {\rm UV},{\rm z=3}}$ results in a value of $1.3\times 10^{-2}$ M$_{\sun}$
yr$^{-1}$ Mpc$^{-3}$ while the integral 
to $10^7$ $L_{\sun}$ yields a SFRD
of 0.04 M$_{\sun}$ yr$^{-1}$ Mpc$^{-3}$. Application of an extinction correction,
inferred to be about $A_{\rm UV}$=0.45 mag at $z\sim6$,
to this latter number, implies a SFRD of 0.06\,M$_{\sun}$~yr$^{-1}$~Mpc$^{-3}$.

The agreement between the SFRD values estimated from ultraviolet
surveys and the GRB rate density is reassuring, considering that there
have been only 8 GRBs that have been spectroscopically confirmed to be
at $z>4$ (Figure 4). However, the SFRD from GRBs primarily
traces the faint end of the galaxy luminosity function while the surveys
are measuring the contribution from the bright end. As a result,
a more reasonable
SFRD estimate requires adding the SFRD contribution estimated from the
faint end of the galaxy luminosity function, from GRBs, to
that from bright LBGs.

GRB hosts are fainter than 0.2 $L_{*,{\rm V, z=3}}$. Based
on the UV to $V-$band flux ratios of star-forming galaxies
at $z\sim3$, it implies that GRB hosts must be fainter than 0.2\,$L_{*,{\rm UV, z=3}}$
Adding the SFRD from $L>0.2~L_{\rm *,UV, z=3}$ galaxies to that inferred
from the GRB rate density results in an extinction corrected
SFRD of $0.27\pm 0.13$ and $0.11\pm 0.05$ M$_{\sun}$ yr$^{-1}$
Mpc$^{-3}$ at $z\sim 4.5$ and $z\sim 6$, respectively. If confirmed through
a larger statistical sample, this is a substantial upward revision suggesting
that $L<0.2~L_{\rm *,UV, z=3}$ galaxies contribute at least four times as much
to the star-formation rate density at $z\sim6$ as the bright end ($L>0.2~L_{\rm *,UV, z=3}$)
of the UV luminosity function. Indirectly, this implies that the faint end slope
of the UV luminosity function at $z\sim6$ must be $\sim-1.9$, compared to the value of
$-1.73$ that was derived by \citet{Bouw:06}.

GRBs are a powerful tool for measuring the high redshift star-formation
rate density.  In particular, deep {\it Spitzer}
observations of GRB hosts can reveal
the contribution to the star-formation rate density from the faint 
end of the galaxy luminosity function,
a regime which is inaccessible to deep, rest-frame ultraviolet/near-infrared
surveys.  Increasing the sample of high redshift GRBs will reduce the
uncertainties in the star-formation rate density unaffected by
extinction and through stacking analysis on the host
galaxies will help estimate the contribution to the stellar mass
density from sub-$L_{*}$ galaxies. Comparison between star-formation rate
estimates from GRBs with those from deep UV surveys will provide
better constraints on the evolution of dust extinction at high
redshift and provide tremendous insights into the chemical enrichment
of the early Universe.

\acknowledgements 

We wish to thank Mark Dickinson for his comments which strengthened 
the arguments in this paper. We also acknowledge the extensive resources
that are invested by the entire GRB community, not all of whom can be cited here,
which enables prompt imaging and spectroscopic
follow-up of the bursts.  This work is based on 
observations made with the {\it Spitzer Space Telescope}, which is operated by the Jet Propulsion Laboratory,
California Institute of Technology under a contract with NASA. Support
for this work was provided by NASA through an award issued by
JPL/Caltech. 
EB acknowledges support by NASA through Hubble
Fellowship grant HST-01171.01 awarded by STSCI, which is operated by
AURA, Inc., for NASA under contract NAS5-26555.

\begin{figure}
\plotone{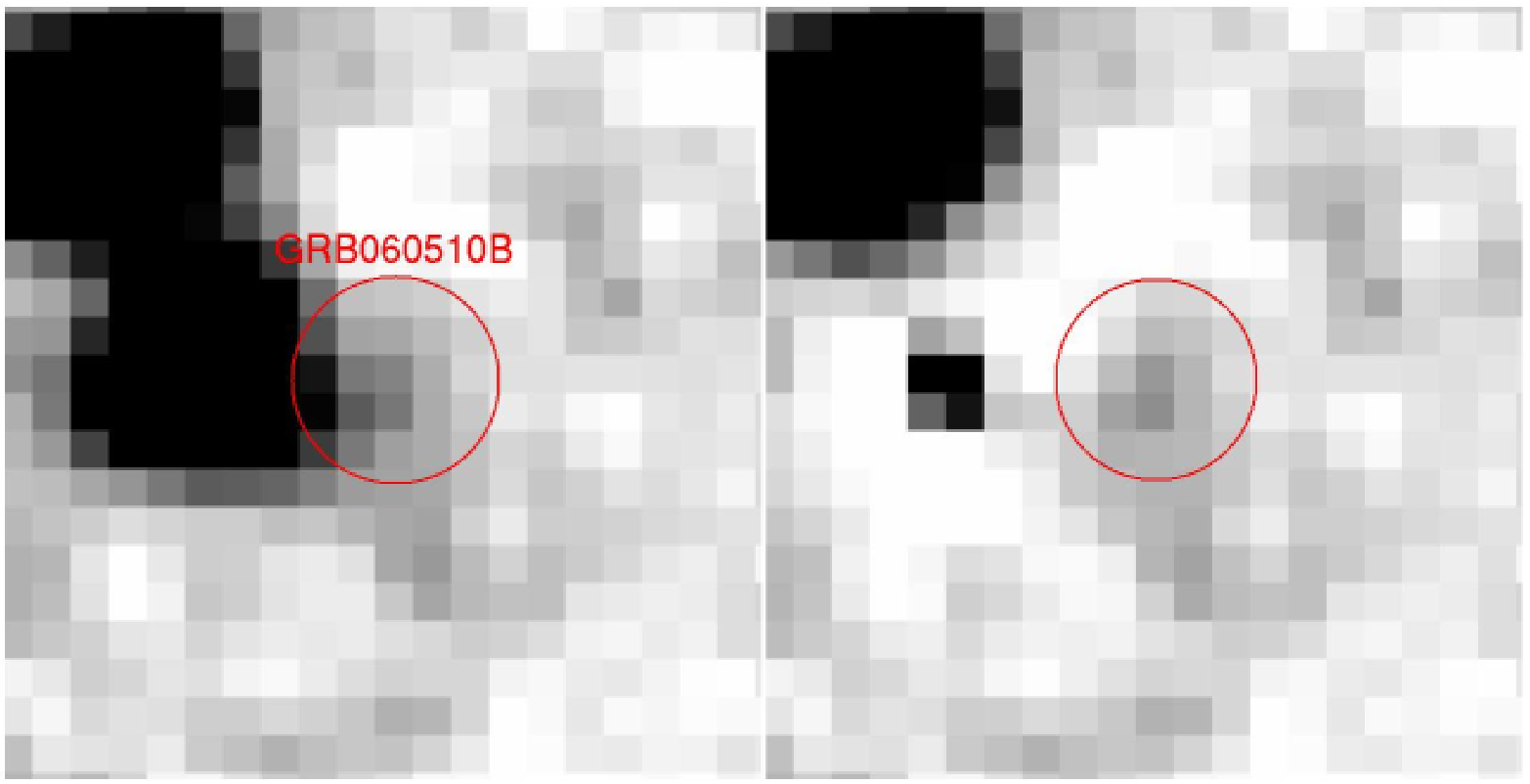}
\caption{\sst\ image of the host galaxy of GRB\,060510B at 
\ra{15}{56}{29.607},\dec{+78}{34}{12.42} (J2000).  Image is 
12$\arcsec$ on a side, North is up, East to the left. The left panel
shows the processed mosaic while the right panel shows the image
with the foreground galaxy 3.1$\arcsec$ to the East
subtracted.  }
\end{figure}

\begin{figure}[hbp]
\plottwo{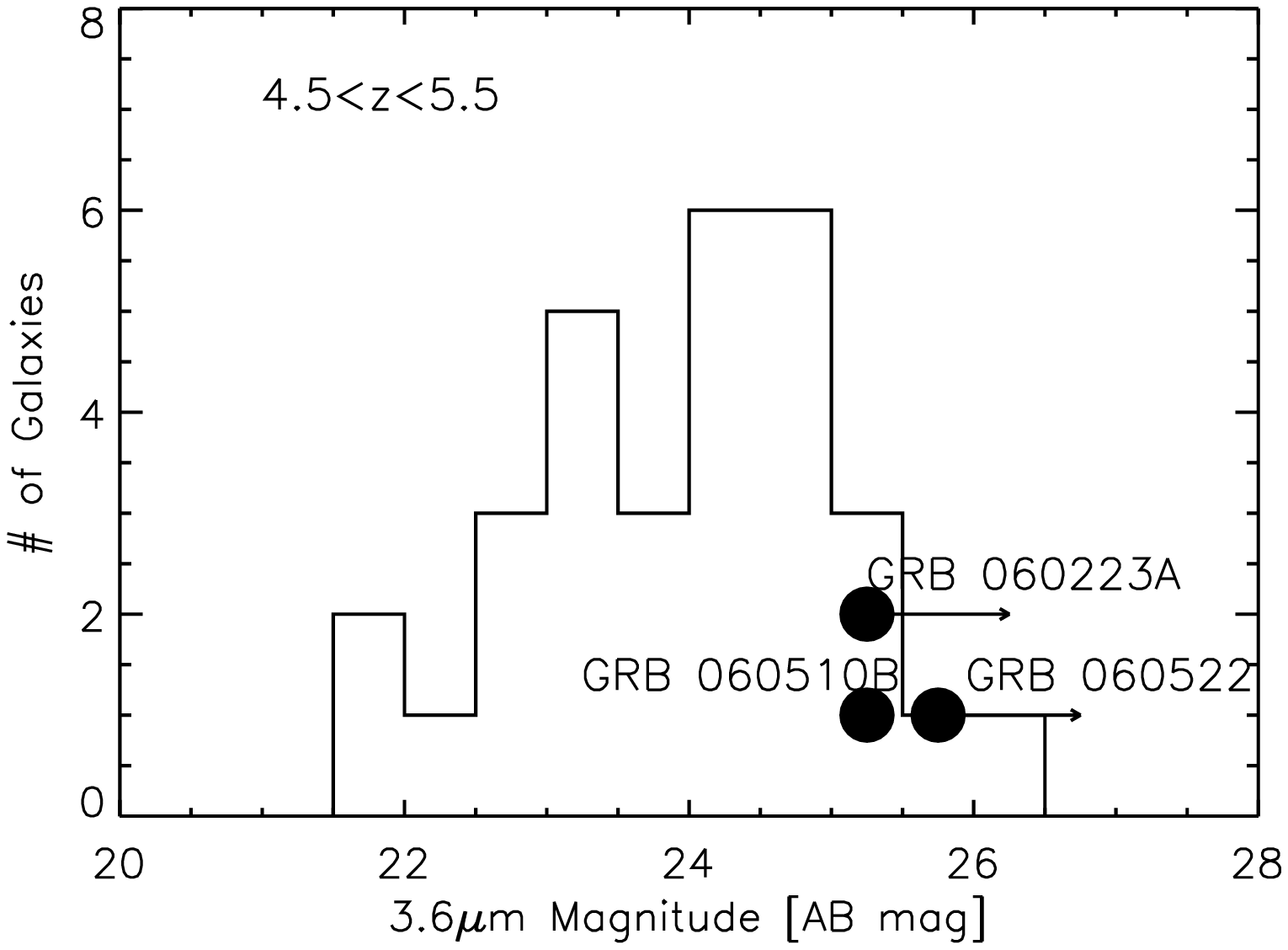}{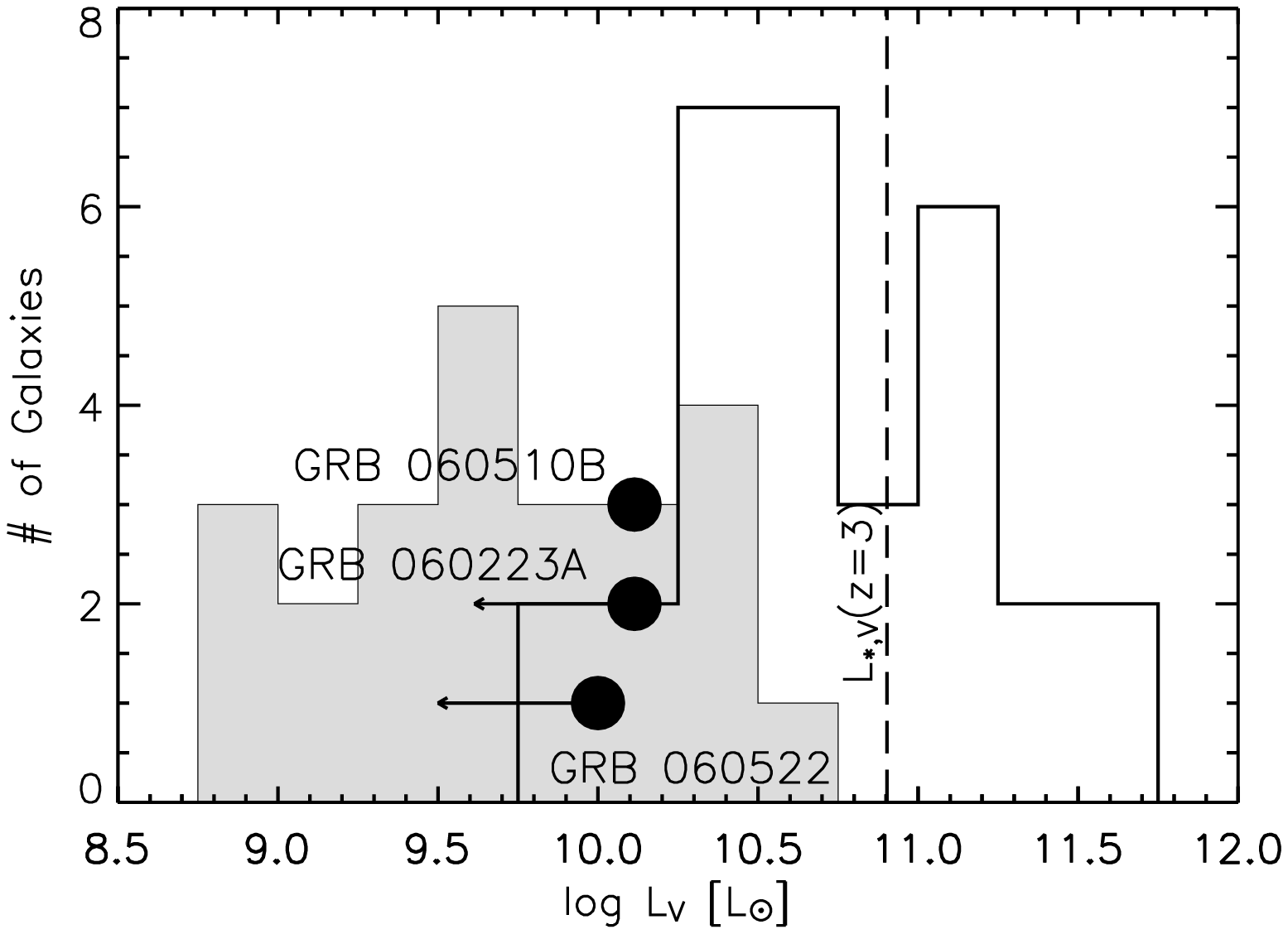}
\caption{(Left) Histogram showing the distribution of observed 3.6 
$\mu$m magnitudes for galaxies in the GOODS fields with spectroscopic
redshifts $4.5<z<5.5$. The solid symbols show the brightness of the GRB host
galaxies observed in this paper relative to the field galaxies.
(Right) GRB hosts have rest-frame $V-$band luminosities which are a
factor of $\sim2-3$ fainter than field galaxies at similar redshifts
and provide a complementary way to study the faint end luminosity
function of star-forming galaxies. Also shown as the shaded histogram is the
$V-$band luminosities of GRB hosts at a median redshift of $\sim$1 \citep{CBA:02, lef:03} which
indicate that GRB hosts span similar $V-$band luminosities, regardless
of redshift.}
\end{figure}

\begin{figure}
\epsscale{0.7}
\plotone{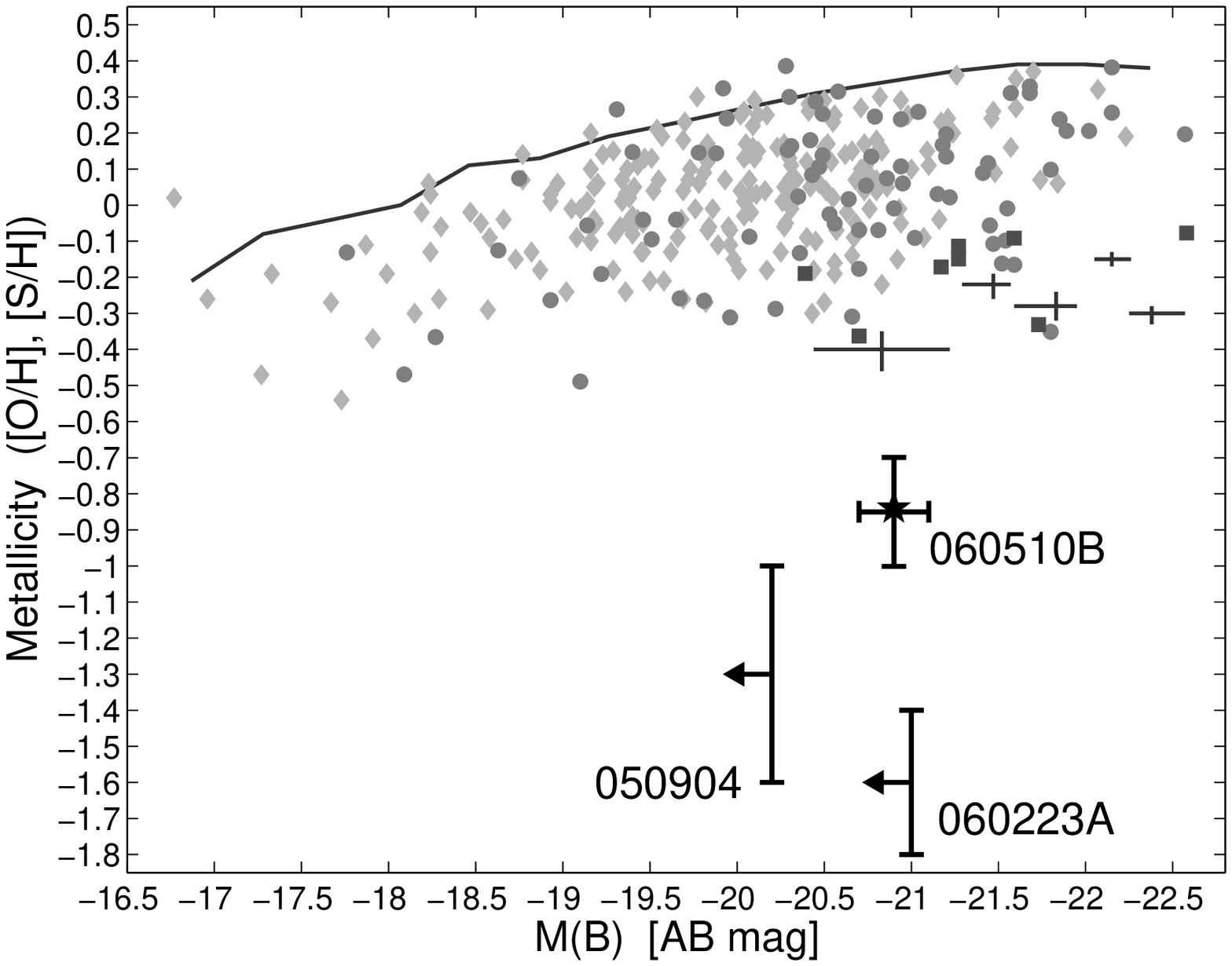}
\caption{
Luminosity-metallicity relationship for star-forming galaxies at $z<2$
compared with the host galaxies of $z>4$ gamma-ray bursts. The galaxy
data are from GDDS and CFRS at $z\sim0.4-1$ 
\citep[circles;][]{Savaglio:05}, TKRS at $z\sim0.3-1$
\citep[diamonds;][]{KK:04}, DEEP2 at $z\sim1-1.5$ 
\citep[squares;][]{Shapley:05} and LBGs at $z\sim2.3$ \citep[error
bars;][]{Erb}. The gray lines represent the relation derived for
$z\sim0.1$ galaxies from the Sloan Digitized Sky Survey
\citep{Tremonti:04}. GRBs provide a unique window into the evolution
of the mass-metallicity relation at high redshift and indicate that
the chemical enrichment of galaxies with redshift occurs at a lower rate than
the build up of stellar mass, presumably due to the expulsion of metals
in low-mass galaxies by outflows.}
\end{figure}

\begin{figure}
\epsscale{0.7}
\plotone{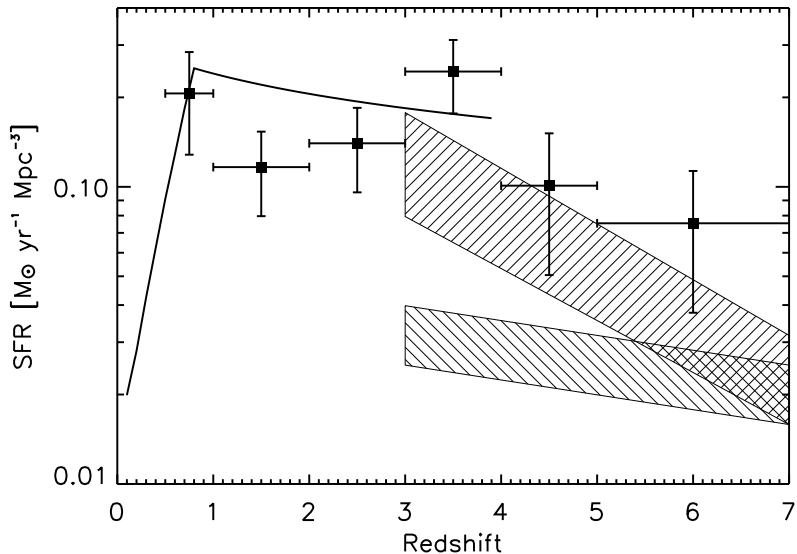}
\caption{Star-formation rate density inferred from spectroscopically 
confirmed long-duration \swift\ GRBs (solid squares).  The solid black
line is the extinction-corrected star-formation rate density inferred
at $z<4$ from a variety of multiwavelength surveys in the mid-infrared
and submillimeter, which are used to calibrate the GRBs
\citep{Chary:01}. The
lower hatched region is the extinction uncorrected SFRD from
rest-frame UV surveys ($>0.04$~L$_{\rm *,UV,z=3}$), including estimates by
\citet{Steidel}, \citet{Yoshida} and \citet{Bouw:06}. The upper hatched rectangle are
these values corrected upward for reddening using the UV-slope
technique by \citet{Bouw:06}. The SFRD inferred from GRBs at $z>4$ is
consistent, within the significant errors, to the extinction corrected
SFRD. Follow-up of a larger number of high-redshift GRBs are required
to confirm if the higher rate density derived from the GRBs is
statistically significant.}
\end{figure}

\begin{deluxetable}{lllccccl}
\tabletypesize{\scriptsize}
\tablecaption{\sst\ Observations of $z\sim5$ GRB Host Galaxies}
\tablewidth{0pt}
\rotate
\tablehead{
\colhead{GRB} &
\colhead{RA, DEC (J2000)} &
\colhead{Redshift} &
\colhead{Date of Observation} &
\colhead{Sky Background\tablenotemark{a}} &
\colhead{Exposure Time} &
\multicolumn{2}{c}{Flux Density in $\mu$Jy} \\
\cline{7-8}
\colhead{} &
\colhead{} &
\colhead{} &
\colhead{} &
\colhead{MJy/sr} &
\colhead{} &
\colhead{3.6$\mu$m} &
\colhead{5.8$\mu$m}
}

\startdata
GRB\,060223A & \ra{03}{40}{49.561},\dec{-17}{07}{48.36} & 4.406 & 2006 Sep 23 & 24.8 & 130$\times$100s & $<$0.3 & $<$2.4\\ 
GRB\,060510B & \ra{15}{56}{29.607},\dec{+78}{34}{12.42} & 4.942 & 2006 Oct 26 & 13.4 & 130$\times$100s & 0.23$\pm$0.04 & $<$2.1 \\
GRB\,060522  & \ra{21}{31}{44.800},\dec{+02}{53}{10.35} & 5.110 & 2006 Nov 23 & 28.4 & 138$\times$100s & $<0.2$ & $<$2.4 \\

\enddata
\tablenotetext{a}{Background at 24$\mu$m, dominated by the zodiacal light.}
\end{deluxetable}
\end{document}